\documentclass[aps,groupedaddress,superscriptaddress,showpacs]{revtex4}
\usepackage{graphicx}
\usepackage{amssymb}
\begin{document}
\title{Coulomb Blockade in a Coupled Nanomechanical Electron Shuttle}
\author{Chulki Kim}
\author{Marta Prada}
\altaffiliation {ICMM-CSIC, Sor Juana Ines de la Cruz, 3, 28049 Madrid, Spain}
\author{Robert H. Blick}
\email{blick@engr.wisc.edu}
\altaffiliation{Electrical $\&$ Computer Engineering, University of Wisconsin-Madison, Madison, Wisconsin 53706, USA}
\affiliation{Physics, University of Wisconsin-Madison, Madison, Wisconsin, 53706, USA}
%
\begin{abstract}
  We demonstrate single electron shuttling through two coupled nanomechanical pendula. 
The pendula are realized as nanopillars etched out of the semiconductor substrate. 
Coulomb blockade is found at room temperature, allowing metrological applications. 
By controlling the mechanical shuttling frequency we are able to validate the different regimes of electron shuttling.
\end{abstract}
\maketitle

One of the fundamental experiments in classical mechanics is the coupled pendula, 
used to demonstrate how energy is transferred from one resonator to another. 
This coupled pendula or coupled oscillator model is often evoked as an analogy for 
the Josephson effect in superconducting junctions, where similar differential 
equations can be applied~\cite{tinkham}. 
In principle, the physics of coupled oscillators can be found in mode-locked 
lasers~\cite{claser}, analyzing weather fronts in climate modeling~\cite{cclimate}, 
and neural networks~\cite{cbio}.

Nanotechnology delivers the tools to fabricate nanoscale electro-mechanical systems, 
pushing towards the ultimate limits of miniaturization~\cite{blick,blencowe,roukes,craighead,cleland}. 
This has already led to the realization of single nanomechanical pendula in 
various forms, {\it e.g.} suspended semiconductor cantilevers~
\cite{erbe_prl,scheible_prl04,koenig_08} and nanopillars~
\cite{scheible_apl04,kim_apl07,kim_njp10}. 
Such a mechanical resonator is commonly placed between two electrodes, so that 
the pendulum can exchange electrons and mechanically transfer them from source 
to drain. Hence, the term electron shuttle was coined~\cite{gorelik,shek}.
Here we present measurements of two coupled pendula realized as nanopillars 
integrated between two contacts. The strongly reduced co-tunneling in two serially 
coupled electron shuttles reveals Coulomb blockade (CB) at room temperature. 
The fundamental importance of CB in nanomechanical shuttles is two-fold: one 
in metrology application as outlined by Weiss and Zwerger~\cite{weiss_zwerger} 
and the second in ultra-sensitive nanomechanical sensors. 

 In classical electron turnstiles~\cite{geerligs}, the stochastic nature of tunneling is typically suppressed by operating the turnstile at frequencies $f$, much lower than the inverse of the time constant $\tau$ (or, more formally, the $1/\tau= 1/RC$-frequency of the contacts). 
In contrast to this, we are able to operate electron shuttles in the regime of high-frequencies ({\it i.e.} $f \sim \tau^{-1}$), and show that clocking of electron transport can be achieved. This effect is due to the suppression of co-tunneling when the shuttles are operated
in series~\cite{weiss_zwerger}. The two shuttles considered here are placed between two nano-scale contacts. We characterize the device's response by probing the direct current through the nanopillars. Mechanical motion is studied first by coupling a DC-bias and then, 
exciting the mechanical motion by adding a radio frequency (RF) signal to the source electrode, as we have demonstrated before~\cite{ckim_prl10}.

 The two nanopillars are defined on a silicon-on-insulator (SOI) substrate, where the top crystalline silicon is $190$~nm thin and the insulating SiO$_2$ is about $350$~nm. A $50$~nm top gold layer serves as the final electrical conduction path. 
The deposited metal is also employed as a mask in a dry etch step, which mills out the SOI material around the pillars. 
 We apply a CF$_4$ plasma etch step and mill into the SiO$_2$ insulating layer,
thus ensuring electron transport {\it via} the metallic islands. Further details on the processing are given by Kim {\it et al.}~\cite{kim_apl07}. In Fig.~1(a), the final sample is shown in a scanning electron microscope graph. The source and drain contacts are placed in close proximity to the two pillars. The distance between the pillars is about 17~nm. The gating electrode is placed further away, enabling a shift in the electrostatic potential of the islands. The inset of Fig.~1(a) gives a broader view of the coplanar-waveguides into which the two nanopillars are embedded. 

 The nanoscale circuit is placed in an impedance matched transmission line in order to minimize signal loss along the line. All measurements are performed under vacuum ($< 10^{-5}$~mbar) in a probe station at room temperature. The station is placed in a Faraday cage and equipped with radio frequency contact probes covering the range from DC to 50~GHz (a bias-tee allows AC/DC superposition with high precision). The equivalent circuit diagram is given in the inset of Fig.~1(b): The two pillars are individually displaced by $x_{1}$ and $x_{2}$, which lead to tunable resistances and mutual capacitances (arrow boxes). The gating electrode couples capacitively to the pillars. The output current of the coupled electron shuttles is fed into a current amplifier.
\begin{widetext}
\begin{figure}[!hbt]
\includegraphics[width=0.8\textwidth]{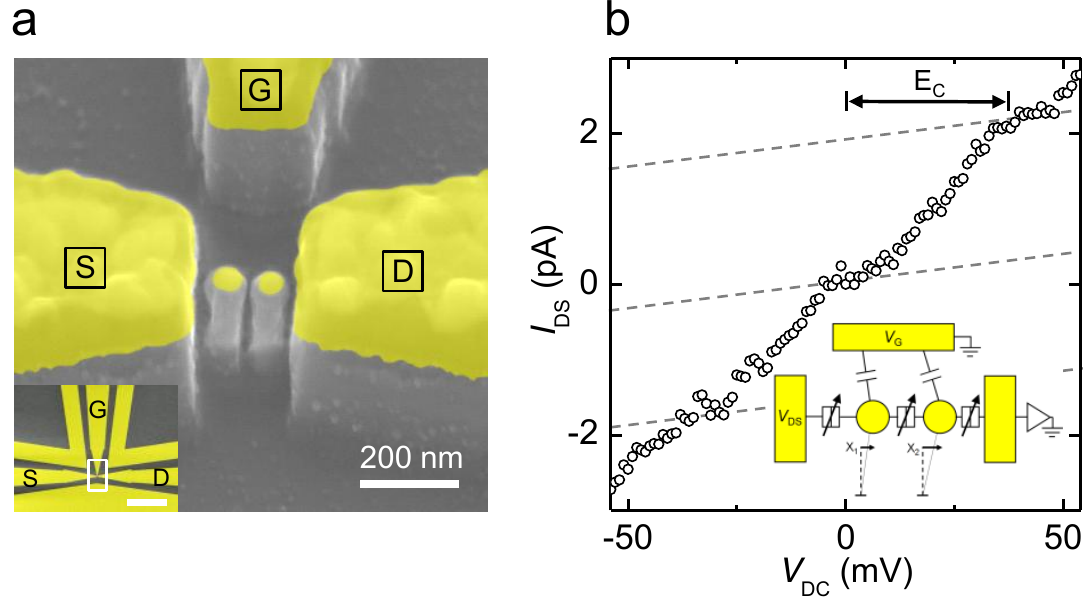}
\caption{(a) Two coupled electron shuttles realized as nanopillars. 
The metallic top layer allows electron exchange with the source ($S$) and drain ($D$) contacts. 
The scale bar corresponds to a length of 200~nm. 
The inset shows a broader view of the coplanar-waveguide into which the nanopillars are embedded. 
The scale bar in the inset is 10~$\mu$m.
(b) DC response: The dashed lines indicate the Coulomb staircase. 
The non-zero slope of the staircase steps is due to thermal broadening. 
The inset gives the circuit diagram with the mechanically tunable tunneling barriers
coupling the leads and shuttles. 
The gate couples capacitively with the two electron islands. 
} 
\label{fig1}
\end{figure}
\end{widetext}

\section{Results and Discussion}
 The resulting $IV$-characteristic is shown in Fig.~1(b): 
the application of a DC bias voltage only across the coupled pendula leads to an ohmic response, modulated by Coulomb blockade steps. The gate electrode is grounded in these measurements. 
The total Coulomb energy $E_{\rm C}$ marks the energy required to transport a single charge through an island~\cite{alex}. 
From the data in Fig.~1(b) (marked by the arrow) we find a total energy of 
$E_{\rm C} \approx 40$~meV. The capacitance of each island of radius $r_{\rm np}$ can be estimated to be $C_{\rm np} \simeq 4 \pi \epsilon_0 r_{\rm np} (1 + (r_{\rm np}/d)^2)$,
where the inter-nanopillar distance is $d = 80$~nm.  
In accordance with the optimized values that fit the experimental model (see {\it Methods} section), 
we obtain $C_{\rm np1} \simeq 3.2$~aF and $C_{\rm np2} \simeq 3.9$~aF, where we used for the islands' radii the 
values $r_{\rm np1} = 25$~nm and $r_{\rm np2} = 31$~nm. Since the two islands are placed in series, the total capacitance is $C^{-1} \approx C_{\rm np1}^{-1}+C_{\rm np2}^{-1} $. 
Hence, we obtain a total capacitance of $C\approx 1.76$~aF, 
from which we extract a Coulomb energy
of $E_C = e^2 /2C = 41$~meV. This is well above the corresponding room temperature energy of 26~meV.
We stress that room temperature CB for a coupled shuttle has not been observed before. 
Conventional single electron shuttles reveal CB only at low temperature~\cite{azuma,azuma2}, 
since suppression of co-tunneling is not as pronounced as for the coupled shuttles considered in this work.

\begin{widetext}
\begin{figure}[!htb]
\includegraphics[width = 0.99\textwidth]{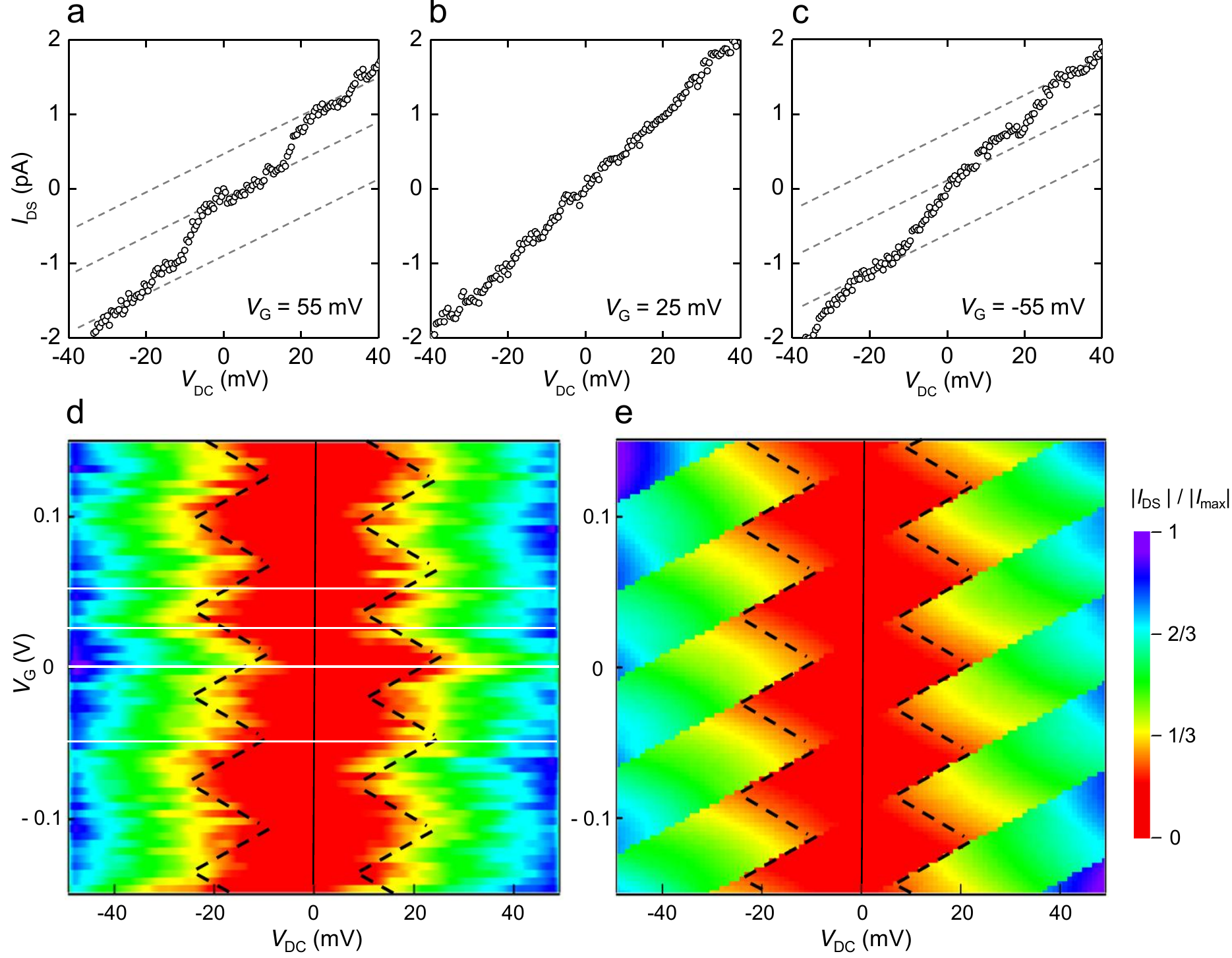}
\caption{Coulomb blockade in the coupled electron shuttle:
(a-c) Current $I_{\rm DS}$ traces {\it vs.} bias voltage $V_{\rm DC}$ for gate voltages 55, 25, and -55~mV.
Experimental (d) and theoretical (e) Coulomb diamonds traced in the normalized current,
$|I_{\rm DS}|/I_{\rm DS}^{\mathrm{max}}$, in color scale representation. 
The lower borders of the CB regions are represented in red with the Coulomb plateaus depicted in green (see color scale). 
The borders of the CB determined from the theory plots are marked by
dashed black lines as a guide to the eye, both in (d) and (e). The theoretically derived lines
trace the measurements closely. The horizontal lines in (d) indicate the previously shown line plots in Fig.~1b and Fig.~2(a--c). 
} 
\label{fig2}
\end{figure}
\end{widetext}


In Fig.~2 the response to a pure DC-bias $V_{\rm DC}$ and gate voltage $V_{\rm G}$ of the coupled shuttles is studied. $IV$-curves at altered gate voltages are shown in Fig.~2(a--c). 
The dashed lines are linear extrapolations of the initial conductance slopes and underline the steps caused by single electron charging. 
Evidently, a typical Coulomb staircase is superimposed on an ohmic response, due to thermal broadening and shuttling~\cite{gorelik}. 
Note that the mechanical motion of the shuttles allow exchange of electrons with one electrode
while suppressing it on other electrode.
Since the applied gate voltage changes the potential around the shuttles, the charging energy required for an additional electron on the 
island can be varied accordingly. The trace at $V_G = 25$~mV in Fig.~2(b) shows an ohmic response without the CB steps. 
As the gate voltage decreases, the Coulomb staircase reappears ($V_{\rm G} = -55$~mV), as shown in Fig.~2(c). 
The traces are not symmetric about zero bias, which is due to the slightly different diameters of the nanopillar islands and 
the possible polarization of an oxide layer on the islands~\cite{wilkins}.
The full DC output signal from a measurement run is summarized with respect to both $V_{\rm G}$ and $V_{\rm DC}$ in Fig.~2(d), 
revealing Coulomb diamonds. We note that the diamond structure compares to that of a single island, due to the common 
$V_{\rm G}$ and the similar size of the islands. The lack of symmetry of the diamonds with respect to $V_{\rm DC}$ = 0 
is due to the actual difference in island radii, and to the difference in capacitance couplings to the source or drain.  
Although the obtained $IV$ curves do not show clear steps due to the thermal environment, the 
periodic variation of the source-drain offset voltages is seen. The line plots for $V_{\rm G} = 55, 25,$ and $-55$~mV are 
indicated by the white horizontal lines in Fig.~2(d).
To validate the experimental results, we compute the current using numerical methods (see the Methods section). 
The theoretical results shown in Fig.~2(e) reproduces the shape and size of the measured Coulomb diamonds
and agrees with the estimated size of the islands. 

\begin{widetext}
\begin{figure}[!hbt]
\includegraphics[width=0.99\textwidth]{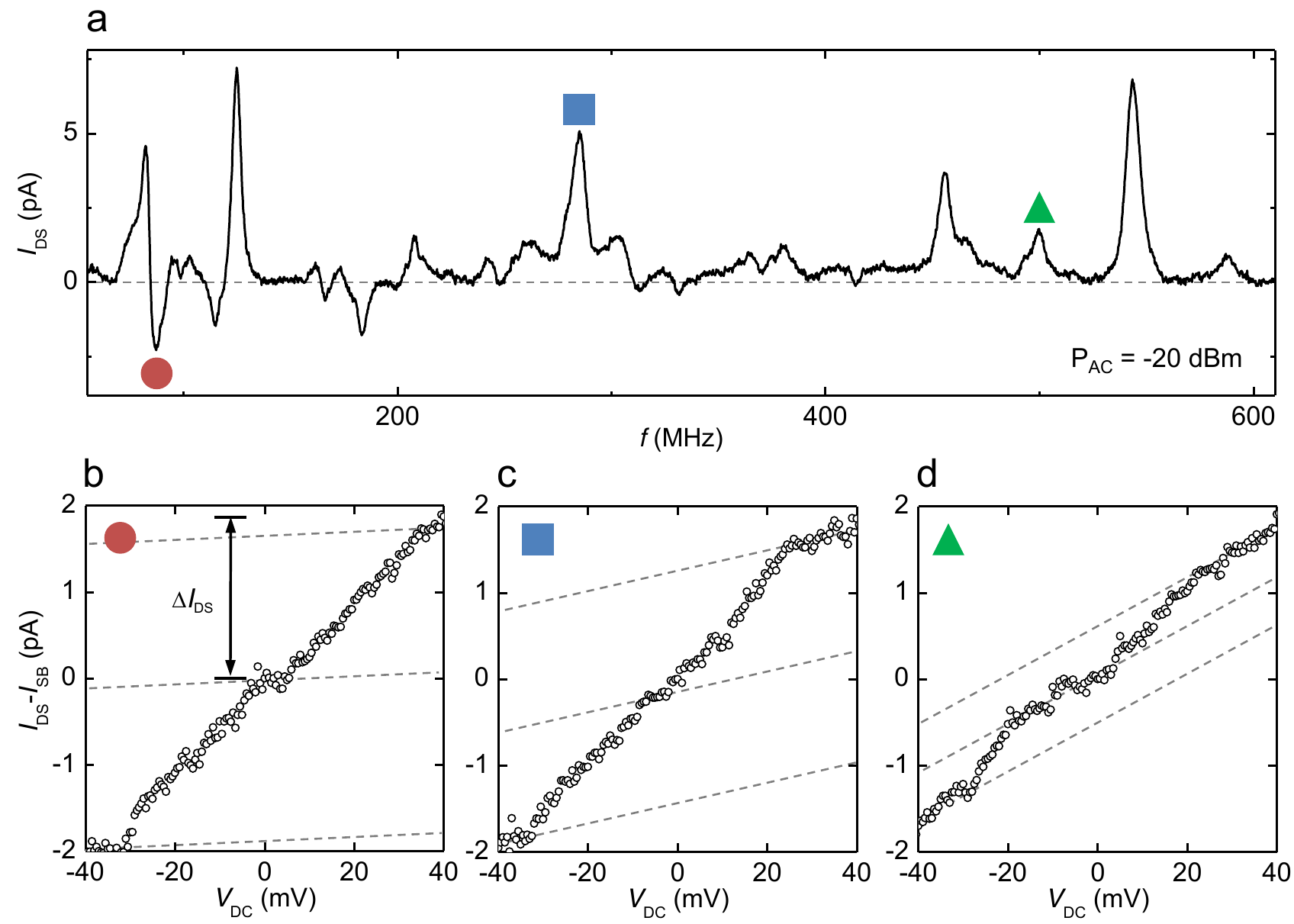}
\caption{
(a) Full frequency sweep of the direct current ($I_{\rm DS}$) through the coupled shuttle revealing the mechanical mode structure for $V_{\rm DC}$ = 0.
The colored symbols indicate three distinct mechanical frequencies plotted in (b-d). 
The lower traces indicate the DC signal at those different modes: $f$ = 87, 285 and 500 MHz, respectively.}
\label{fig3}
\end{figure}
\end{widetext}


 Next we apply an RF signal with zero DC bias in order to tune the mechanical 
motion of the system, as shown in Fig.~3(a). The ability to `dial in' the shuttling frequency is the fundamentally new feature of this device, as compared to classical single electron transistors. 
We can now trace the DC bias voltages at the three different mechanical modes marked in Fig.~3(a), 
plotted in Fig.~3(b), (c) and (d), respectively.
These correspond to commensurate mode numbers $p/q$ of the fundamental mode (for more details, see~\cite{ckim_prl10}): 
$p/q$ is 1/6 (circle, $f$ = 87MHz), 5/9 (square, $f$ = 289MHz) and 1/1 (triangle, $f$ = 500MHz).
The current offset $I_{\rm SB}$ due to spontaneous symmetry breaking (SB) 
in coupled electron shuttles~\cite{ckim_prl10} is subtracted.  
Dashed lines extend the inclined plateaus to underline the staircase current.

\begin{widetext}
\begin{figure}[!htb]
\includegraphics[width=0.95\textwidth]{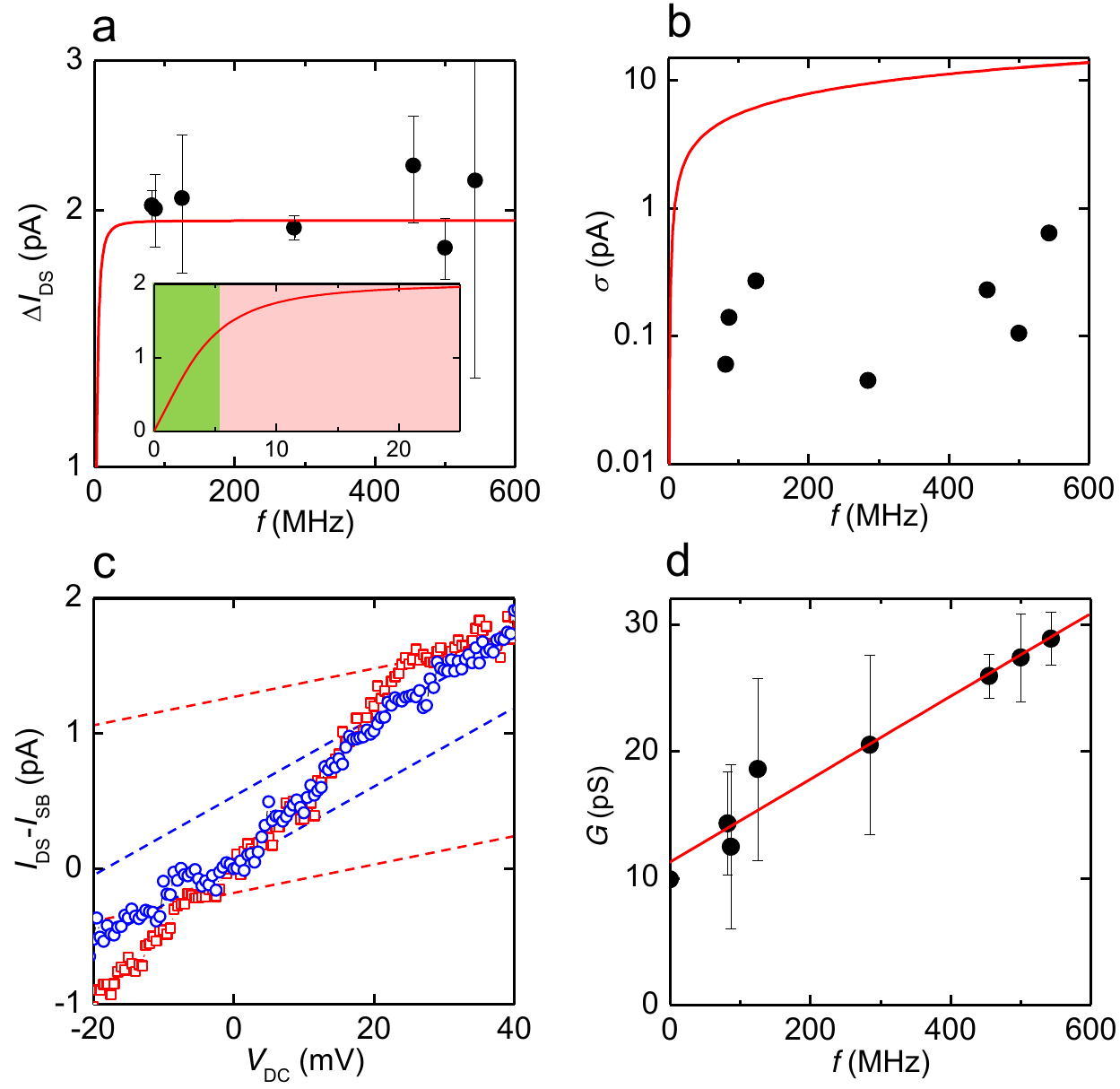}
\caption{
Evaluation of time dependent shuttling:
(a) The current steps between plateaus $\Delta I_{\rm DS}$ are plotted {\it vs.} the 
mechanical frequency $f$ at a fixed gate voltage ($V_{\rm G}$ = 0~V). 
The measured current steps (black dots) fall on the expected flat-line, as predicted~\cite{weiss_zwerger}. 
The inset shows a theoretical plot of the expected current in the low frequency region. The coupled shuttles are operated above the cutoff frequency $\tau^{-1}$ (red region).
(b) The standard deviation $\sigma$ of the measured current (black dots). 
This is compared to the theoretical expression for charge fluctuation in a single shuttle (red solid line) as obtained from ~\cite{weiss_zwerger}. The strongly reduced co-tunneling leads to a suppression of fluctuations, allowing the observation of CB. 
(c) $IV$-traces for two mechanical modes, as seen the slopes in the plateau increase from a shuttling frequency of 285~MHz (red squares) to 500~MHz (blue circles).
(d) Summary of the differential conductance within the Coulomb-plateau for the different shuttling modes. As expected the effective differential conductance is increased to the mechanically enhanced escape rate of the electrons.} 
\label{fig4}
\end{figure}
\end{widetext}

Fig.~4(a) exemplifies the dissipative regime: In this limit, a 
saturation current (frequency independent) is achieved, as the effective contact time of 
the islands with the electrodes is proportional to the inverse of the frequency, and at 
the same time, the current is proportional to the time contact and to the frequency 
(see {\it Methods} section). 
We observe a saturation value for the current of about 2~pA.   
The measured current steps (black dots) fall on the expected 
shuttle current (red solid line), as predicted from \ref{eq:Nav}. 
We have to note that their calculation only considered a single shuttle.
The inset in Fig.~4(a) shows a magnified view 
of the expected current in the low frequency region. 
The coupled shuttles are operated in the high frequency, dissipative limit (red region).
We estimated $x_{\rm max}\sim$~5-8 nm, $d\sim$~17 nm, and $\lambda\sim$ 1~nm, giving $t_0\sim 1/(2\omega)$.
We also have $\tau\sim\times10^{-7}$s. Thus, we find the transition frequency to be of the order of 
5 MHz, which agrees very well with our data.  
The standard deviation $\sigma$ of the experimental values are found to be much 
lower than the theoretically expected fluctuations, as seen in Fig.~4(b).
This can be ascribed to the fact that the model assumed a single shuttle 
unlike the actual device of two coupled charge shuttles in series. 
However, charge fluctuations are even more strongly suppressed off by the mechanical resonances for two coupled shuttles. 

We consider now the conductance level within the plateau. 
Fig.~4(c) shows  two $IV$-traces at 285~MHz (red squares) and 500~MHz (blue circles). 
The dashed lines give the slopes of the corresponding plateaus in the neighboring Coulomb regions. 
The slope, or differential conductance $G= \Delta I_{\rm DS}/\Delta V_{\rm DC}$, increases linearly
towards higher shuttling frequencies, a consequence of self-excitation. 
This occurs around $V_{\rm DC}$ = 0, where a mechanical mode was 
located, and at $V_{\rm DC} > V_c$. 
In the CB regime, the number of transferred electrons per period is quantized in units of $e$, 
$Ne\propto (V_{\mathrm{DC}} C /e + 1/2)e$. 
Moreover, in the self-oscillatory regime (above a critical voltage, $V_c$ \cite{gorelik}, 
or around $V_{\rm DC}=0$, where the mechanical mode was excited), the charge 
exchange with the leads takes a fixed quantized value corresponding to the thermal equilibrium with the 
nearest lead, and is exponentially suppressed with the far lead. 
Following Gorelik {\it et al.} we estimate the critical voltage $V_c$ to be of the order of the CB
voltage, $V_c\sim e/C$ \cite{gorelik}. 
In this regime, the conventional linear-in-frequency relation applies, $I\simeq 2Nef$.
Note that for intermediate $V_{\rm DC}$ values, 0$<V_{\rm DC}<$
$V_{c}$, the amplitude of the self-oscillations is suppressed. In this region, \ref{eq:Nav} applies, 
rather than the linear-in-frequency relation (see also the {\it Methods} section). 
The overall results are summarized in Fig.~4(d), where we show the step differential conductance in the 
plateau, $\Delta I_{\rm DS}/\Delta V_{\rm DC}$, or the slope of the dashed lines of Fig.~4(c). 
The linearity in frequency of the
plateau differential conductance $G$ is evident. This demonstrates the self-excitation 
of the nanopillars, already shown to cause ohmic behavior~\cite{kim_njp10,zettl_nl10}.

\section{Conclusion}
  In summary we observe Coulomb blockade in a coupled electron shuttle operating at room temperature. 
Mechanical clocking of electron transport has been demonstrated by using RF signals.  
Operating in the high frequency regime leads to a strong suppression of current fluctuations.
Two regimes of operation are observed in the high frequency limit as a function of bias voltage, 
corroborating theoretical predictions: the dissipative regime, where the current is frequency independent,  
and the self oscillatory CB regime, showing a linear-in-frequency conductance. 

\section{Methods}
\label{methodsec}
 To validate the experimental results obtained in Fig.~2(a-d), we compute the current within the orthodox model of CB. 
The accessible states are described by their respective probabilities $P_{n_1,n_2}$, where $n_i$ is the 
number of excess electrons on the islands $i=1,2$. 
In the absence of an RF signal, the system can be assumed to be in the dissipative regime, 
where the amplitude of the mechanical oscillations is small and the tunneling current prevails 
over the shuttle current \cite{gorelik}. 
Changes in the resistance of a particular junction are proportional to $e^{x_i/\lambda}$, 
with $\lambda$ being the typical tunneling distance, $x_i$ the displacement, 
and $x_i\lesssim\lambda$ in the dissipative regime. 
In this limit, we suppose 1/$\Gamma$ to be much smaller than the typical time scale of the displacements, 
and the time evolution can be described in terms of coupled master equations~\cite{ahn}, which we solve by 
exact diagonalization.   
In the CB limit, we observe that 
$\langle |I_{\rm DS}|\rangle/I_{\rm DS}^{\mathrm{max}}\propto \sum_{n_1,n_2} P_{n_1,n_2} 
(\overrightarrow{\Gamma}_{n_1,n_2} - \overleftarrow{\Gamma}_{n_1,n_2}) $, 
where $\langle\dots\rangle$ denotes an ensemble average. $\overrightarrow{\Gamma}_{n_1,n_2}$ 
gives the rate of tunneling from left to right at one of the junctions for a given configuration $\{n_1,n_2\}$: 
\begin{equation}
\overrightarrow{\Gamma}_{n_1,n_2} =\left( \frac{1}{e^2R}\right)
\frac{\Delta E(n_1,n_2)}{\exp{[\Delta E(n_1,n_2)/k_BT] } - 1 }, 
\label{eq:gamma} 
\end{equation}
with $R$ being the average resistance of the junction and $\Delta E$ the chemical potential.  
The theoretical results are shown in Fig.~2(e): 
the CB region below the first transition is shown in red. 
We stress that a finite conductance is observed in this region due to thermal broadening and direct electron shuttling. 
In the dissipative regime, however, we assume temperature broadening to be the dominant source of the ohmic response. 

We now focus on the ohmic response between the differential conductance plateau or 
current step, $\Delta I_{\mathrm {DS}}$. 
Conventionally, one would expect a simple linear relation between this ohmic current 
and the shuttling frequency of the form, $I = 2ef$ or $\Delta I_{\mathrm {DS}} \sim f$, which 
is typically applied for electrometry purposes~\cite{nato_asi}. 
This relation, however, is only valid in the limit of large tunneling rates, 
$\overrightarrow\Gamma_{i,j} \gg f^{-1}$ with $\overrightarrow\Gamma_{i,j}$ defined in \ref{eq:gamma}. 
For our setup, the relation applies only in the self-oscillatory regime, achieved in the plateau or around 
$V_{\rm DC}=0$, where a mechanical mode is excited. 
Our theoretical results indicate that in the inter-plateau region $ \Gamma_{i,j}\lesssim$~5-15~MHz, which is slower 
than the mechanical motion. Hence, we can state that the 
coupled shuttles operates in the dissipative regime. According to Weiss \& Zwerger~\cite{weiss_zwerger}, we can calculate in this regime 
the average number of transferred electrons $\langle N \rangle$ per period: 
\begin{equation}
\langle N \rangle =  \frac{2(1-a^3)}{(1+a)(1+1/2 a+a^2)},
\label{eq:Nav}
\end{equation}
 where $a \equiv \exp(-t_0/\tau)$, and 2$t_0$ is the effective contact time, 
$t_0 \equiv \omega^{-1} \sqrt{\frac{\pi \lambda}{2 x_{\rm max}}}  (1+ \frac{\lambda}{2d})$. 
The length $x_{\rm max}$ indicates the maximum displacement of the mechanical oscillation, 
and $\omega = 2 \pi f$.
In the regime of high RF frequencies, the effective contact time is shorter than the charge relaxation time, $t_0\ll \tau$.  
Thus the number of transferred electrons per period is inversely proportional to the frequency of the shuttle oscillation,  
$\langle N \rangle \propto t_0$ 
with $t_0 \propto 1/f$. 
Then, the electron shuttle current achieves a frequency independent 
saturation value, $I = \langle N \rangle e f \propto \Delta I_{\rm DS}$.

{\bf{Acknowledgements}}
We are grateful to G. Platero for enlightening discussions. 
The authors like to thank DARPA for support through the NEMS-CMOS program (N66001-07-1-2046), 
the University of Wisconsin-Madison for support with a Draper-TIF award and the Spanish 
Ministry of Education, program (SB2009-0071).




\end{document}